\documentclass[twocolumn,pra,aps,amsmath]{revtex4}

\usepackage{graphicx}

\begin{document}

\title{An electron Talbot interferometer}
\author{Benjamin J. McMorran}
\email{mcmorran@physics.arizona.edu}
\author{Alexander D. Cronin}
\affiliation{Department of Physics, University of Arizona, Tucson, AZ 85721}

\begin{abstract}
The Talbot effect \cite{TAL36}, in which a wave imprinted with transverse periodicity reconstructs itself at regular intervals, is a diffraction phenomenon that occurs in many physical systems. Here we present the first observation of the Talbot effect for electron de Broglie waves behind a nanofabricated transmission grating. This was thought to be difficult because of Coulomb interactions between electrons and nanostructure gratings, yet we were able to map out the entire near-field interference pattern, the ``Talbot carpet", behind a grating. We did this using a Talbot interferometer, in which Talbot interference fringes from one grating are moir\'{e}-filtered by a 2nd grating. This arrangement has served for optical \cite{PAT89}, X-ray \cite{MKS03}, and atom interferometry \cite{CEP95}, but never before for electrons. Talbot interferometers are particularly sensitive to distortions of the incident wavefronts, and to  illustrate this we used our Talbot interferometer to measure the wavefront curvature of a weakly focused electron beam. Here we report how this wavefront curvature demagnified the Talbot revivals, and we discuss applications for electron Talbot interferometers.
\end{abstract}

\date{\today}
\maketitle

Electron optics is a highly developed field, but Talbot interferometry with electrons is new. In transmission electron microscopy Talbot revivals (Fourier self-images) behind crystals have been imaged directly, and understanding these revivals is necessary for the correct interpretation of crystal strucuture \cite{SPE03}. However, direct images of Talbot revivals are not nearly as sensitive to wavefront distortions as the signal from a Talbot interferometer.  As we discuss, our arrangement of two nanogratings can easily detect a beam convergence of $10^{-4}$ radians. Nanogratings have been used recently to construct other types of electron interferometers - a Lau type \cite{CRM06} and a Mach-Zehnder type \cite{GBB06} - but both of these designs are insensitive to wavefront deformations in the incident electron beam.

Observations of the Talbot effect with atoms \cite{CEP95,NKD97} launched many applications for near-field atom optics, such as compound beam splitters for atomic de Broglie waves \cite{WAW05,GDS06}, Talbot-Lau interferometers for atoms and large molecules \cite{CLL94,BHZ02}, interferometry with the Poisson spot for atom waves \cite{NSM98}, and ``direct deposit" lithography of atoms behind phase and absorption gratings \cite{TBB92,MSC93}. Talbot interferometers have found numerous applications in light optics too, such as imaging phase objects \cite{LOS71}, measuring beam collimation \cite{SIL71}, and characterizing lenses \cite{BHA91}. For a review see \cite{PAT89}. Recently, nanostructures have been used to build X-ray Talbot interferometers \cite{MKS03,WCZ05}, which provide images of phase objects with less X-ray dose delivered to the subject. Imaging phase objects is possible because the Talbot interferometer is a type of shearing interferometer \cite{BAT47,RON64}, in which a beam is split into multiple overlapping paths. These applications, which which have been realized with atoms or photons, suggest potential uses for \emph{electron} Talbot interferometers.

\begin{figure}
\includegraphics[width = 8cm]{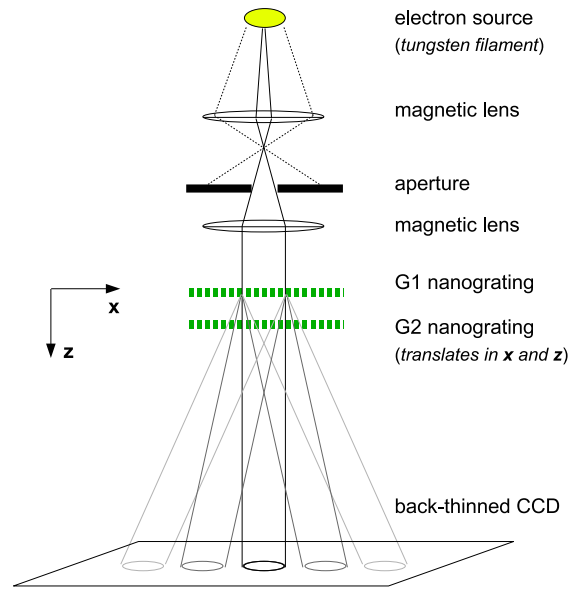}
\caption{\textbf{Experimental setup.} A beam of electrons illuminates a nanoscale transmission grating G1 (at location $z$ = 0 along the optical axis). The resulting near-field interference pattern is read out using a second identical grating G2 (at $z$ = 0.1-1.7 mm). An imaging detector (at $z$ = 1 m) records the transmitted far-field intensity distribution.}
\label{fig:setup}
\end{figure}

A diagram of our Talbot interferometer is shown in Figure \ref{fig:setup}. The nanofabricated gratings have a period $d=$ 100 nm and serve as Ronchi rulings for low energy ($<$10 keV) electrons \cite{MPC06}. When grating G1 is illuminated by collimated plane waves with wavelength $\lambda$, Fourier images (Talbot revivals) of the grating occur at half-integer multiples of the Talbot distance $L_T = 2d^2/\lambda$. For 2.8 keV electrons, $\lambda =$ 23 pm and $L_T =$ 0.86 mm. The spatial modulations of these images, 100 nm in this case, are too small to resolve with our imaging detector, but they can be analyzed using a second grating G2.

\begin{figure}
\includegraphics[width=8cm]{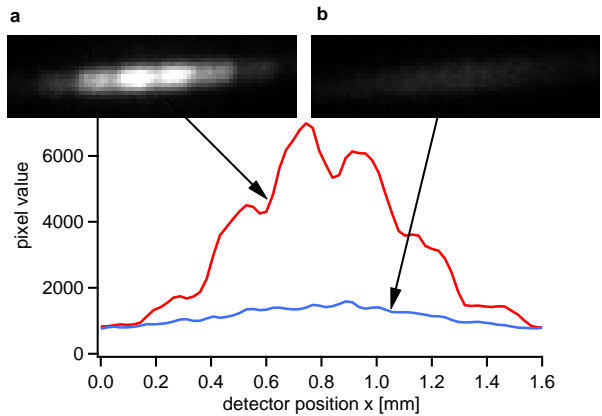}
\caption{\textbf{Far-field diffraction of electrons transmitted through two nanogratings} using a 2.8 keV collimated beam. The gratings were separated by one Talbot distance $L_T$ = 0.86 mm. In (a) the slits of G2 are in registry with the Talbot fringes of G1, and in (b) G2 is shifted in the $x$-direction by 50 nm, half a grating period. Line profiles of each image are indicated below them. The asymmetry of diffraction orders (barely resolved due to the wide, collimated beam) is discussed in the text.}
\label{fig:tal}
\end{figure}

The total flux of collimated electrons transmitted to the far-field is maximum when the slits of G2 line up with the fringes of a Talbot revival from G1, as shown in Figure \ref{fig:tal}(a). In Figure \ref{fig:tal}(b) the transmitted flux is reduced by 77\% when G2 is shifted laterally by half a period, since the Talbot fringes are then blocked by G2's grating bars. This modulation of the total transmitted intensity (shown in Figure \ref{fig:tal}) occurs only when the gratings are illuminated by the plane waves of a well-collimated beam.

\begin{figure*}
\centering
\includegraphics[width = 17cm]{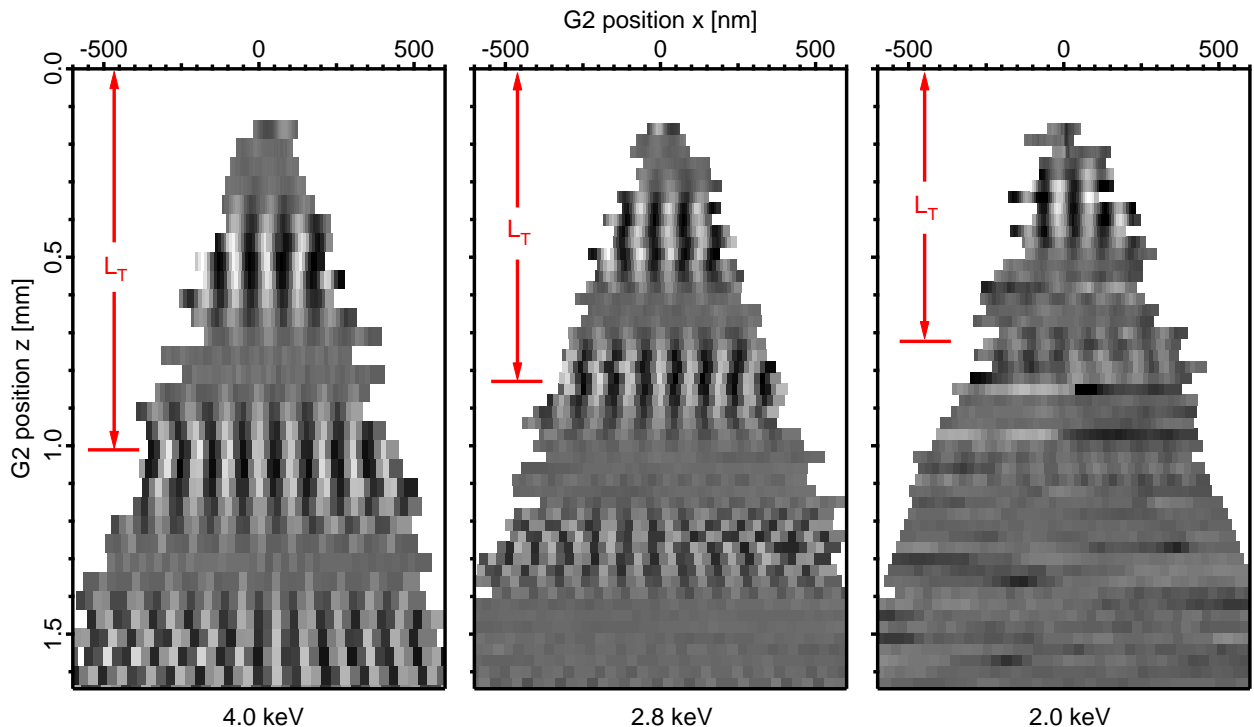}
\caption{\textbf{Electron near-field interference, or `Talbot carpets', behind a nanograting,} using electrons with energy 4 keV (left), 2.8 keV (center), and 2.0 keV (right). The Talbot distance is indicated for each energy (de Broglie wavelength). The value of each pixel is proportional to the total integrated intensity in the far-field diffraction pattern, detected for a particular position of G2 relative to G1. Distortions in these images are due to experimental uncertainty in the position of G2, and are not attributable to distortions in the incident electron waves.}
\label{fig:carpets}
\end{figure*}

Images of the entire electron near-field interference pattern behind nanograting G1, also known as Talbot carpets, are shown in Figure \ref{fig:carpets}. These data were obtained by scanning the position of analyzing grating G2 throughout the near-field region of G1, in both the $x$ and $z$ directions, while recording the total transmitted electron intensity. The tapered shape of each image is due to the limited lateral scan range at small G1-G2 separations.

\begin{figure}
\centering
\includegraphics[width = 8.5cm]{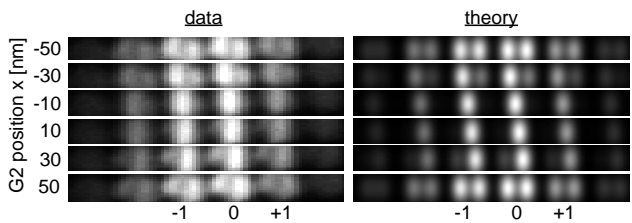}
\caption{\textbf{The demagnified Talbot effect, indicated by far-field diffraction patterns collected at different lateral positions of G2.} The left column is a series of diffraction patterns, with the diffraction orders indicated below them, recorded when using a weakly focused beam of 2 keV electrons. An intensity null moves sideways through each diffraction order as G2 is moved laterally. The simulations in the right column reproduce this modulation, using a model \cite{MCC08} in which G2 filters the demagnified Talbot revival from G1.}
\label{fig:demagtal}
\end{figure}

\begin{figure}
\centering
\includegraphics[width = 8cm]{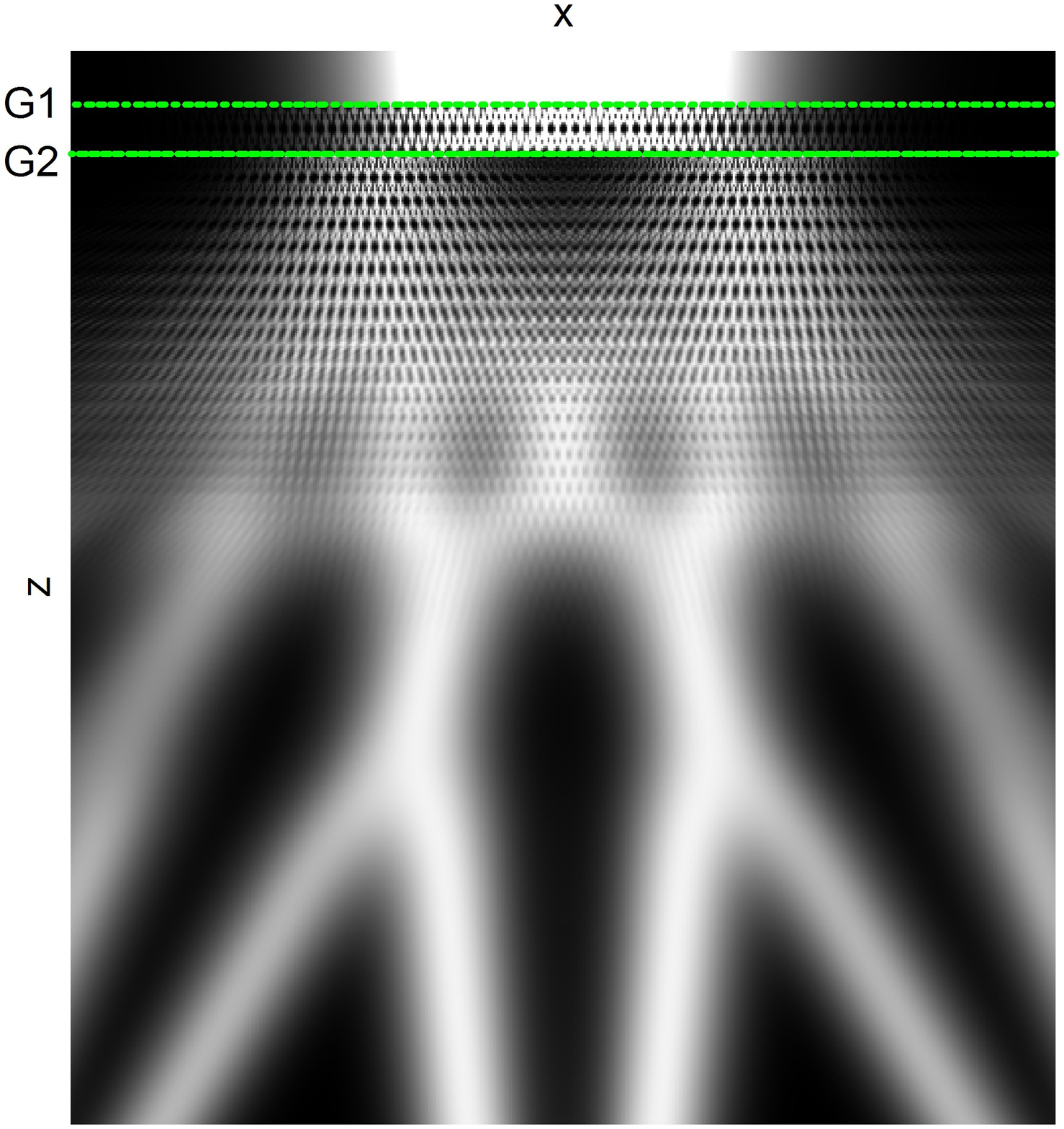}
\caption{\textbf{Simulation of a Talbot interferometer with a converging incident beam.} The Talbot revival has a smaller period than the reference grating. Only the edges of the converging beam are transmitted, which results in two tilted far-field diffraction patterns. This simulation was made using the model developed in \cite{MCC08}.}
\label{fig:focusedtalsim}
\end{figure}

When G1 is illuminated with a converging beam, a different type of modulation is observed (Figure \ref{fig:demagtal}). In this condition, the Talbot revivals from G1 have a finer spatial period than the reference grating G2, due to geometrical demagnification by converging spherical wavefronts. As illustrated in a simulation shown in Figure \ref{fig:focusedtalsim}, G2 blocks some parts of the beam but not others; i.e. there is a moir\'{e} effect between G2 and the interference pattern from G1. This causes dark spots to appear within the resolved far-field diffraction orders. When G2 is scanned laterally, these dark nulls move sideways through the diffraction pattern, as shown in Figure \ref{fig:demagtal}. This behavior is well-described by a general theoretical model that we developed for grating interferometers \cite{MCC08}. The only free parameter in this simulation was the radius of wavefront curvature of the incident beam. The theory (right column in Figure \ref{fig:demagtal}) matches the data (left column) best using an incident radius of wavefront curvature equal to $2.15\pm0.1$ m, corresponding to a nearly-collimated beam with a convergence angle of $\sim$75 $\mu$rad. Negative diffraction orders in Figure \ref{fig:demagtal} have a higher intensity than their positive counterparts, and this asymmetry is understood to result from image-charge (Coulomb) interactions between the grating and transmitted electrons \cite{MPC06}.

This work shows that the Talbot effect is a way to reproduce periodic structures using electron beams. Similar gratings have been used as masks for projection electron lithography \cite{YOS02}. However, deliberate use of the Talbot effect for lithography would be advantageous because the mask could be located millimeters from the substrate, since a Talbot revival, not a direct shadow, would be used for the exposure. Furthermore, as we have shown here, demagnified Fourier images created using a focused electron beam could be used to construct structures with finer periods than the original (a similar technique has already been demonstrated using UV lithography \cite{YMP99}).

To summarize, we have built a Talbot interferometer for 2.8 keV electrons using two nanofabricated gratings. We used this device to map the near-field interference pattern, known as the Talbot carpet, behind a single grating. Analogous to X-ray and optical Talbot interferometers, this arrangement is very sensitive to deformations in the wavefronts of incoming electrons. We demonstrated this by measuring the 2.1-meter radius of wavefront curvature of a focused electron beam, and creating demagnified Talbot revivals with features smaller than the original grating. Both the imaging and lithographic capabilities afforded by the scaled Talbot effect will be explored in future work.

\section{Methods}
The gratings were made by the NanoStructures Laboratory at MIT using achromatic UV interferometric lithography \cite{SSS96}, and are notable for their long range spatial coherence over a large (0.5 $\times$ 5 mm) area. They consist of an array of 50-nm-wide slits etched all the way through a 150-nm-thick suspended membrane of low stress Si{$_3$}N{$_5$}. The distance between adjacent slits, the grating period, is 100 nm. To enable their use as a diffractive optic for electrons, the gratings were sputter-coated with approximately 4 nm of Pt to prevent charging. Both gratings were identically prepared in the same batch process.

The electron beam was provided by a standard SEM electron optics column \cite{GNJ03} featuring a tungsten hairpin filament source. The column was adjusted to produce an approximately collimated, $\sim${\O}150 $\mu$m electron beam. The two gratings were mutually aligned about the optical ($z$) axis of the beam using a motorized rotation stage. The $z$ separation distance between the gratings was adjusted in 30 $\mu$m steps from 0.1 to 1.7 mm using a motorized translation stage. For each separation distance, the gratings were shifted laterally (in the $x$ direction) with respect to the each other by tilting the mount that held them about an axis parallel to the gratings bars. An imaging detector placed 1 m downstream from the gratings collected images of the transmitted electrons, shown in Figures \ref{fig:tal} and \ref{fig:demagtal}. To create the converging beam for the demagnified Talbot effect (Figure \ref{fig:demagtal}), a magnetic lens before the gratings was used to weakly focus the electron beam.

The electron detector is novel. A Princeton Instruments PIXIS-XO camera designed for X-ray and EUV imaging was used to directly image low energy (0.3-5 keV) electrons with high sensitivity. Back-thinned, back-illuminated CCDs have been used to directly detect low energy electrons before \cite{HOR03}, but to our knowledge this work is the first time a commercially-available camera has been used to directly image electrons.

Near-field interference fringes were revealed by moving G2 with respect to G1 in the $x$ and $z$ directions.  The total transmitted electron flux was determined for each position of G2 by
summing the value of all the pixels in images such as Figure \ref{fig:tal}(a).  In total, 5200 images were acquired over a period of 3 hours (only 260 seconds of beam time was needed) to create the Talbot carpets in Figure \ref{fig:carpets}. Each row in Figure \ref{fig:carpets}, which corresponds to a particular grating separation, was shifted after the data were acquired so that intensity peaks in adjacent data sets lined up.  This was necessary because small displacements (of order 50 nm) in $x$ were unavoidable while shifting the $z$ separation by hundreds of microns.

The simulations of the demagnified Talbot effect (in the right column of Figure \ref{fig:demagtal} and in Figure \ref{fig:focusedtalsim}) use a diffractive optical theory developed in \cite{MCC08} based on Gaussian Schell-model beams. Input parameters to the simulation in Figure \ref{fig:demagtal} - such as the initial beam width, spatial coherence, and electrostatic interactions between transmitted electrons and the grating - were obtained in previous measurements \cite{MPC06}, and only the radius of wavefront curvature was left as a free parameter. The parameters used for the simulation in Figure \ref{fig:focusedtalsim} are not based on experiment - they were chosen in order to best illustrate the various interference phenomenon throughout the Talbot interferometer under focused illumination.

\section{Acknowledgements}
We acknowledge Manjul Shah at Princeton Instruments for the loan of the PIXIS imaging detector, Tim Savas at MIT for fabricating the gratings, and Mark Robertson-Tessi and Grady Weyenberg for helping to construct the electron beam apparatus. This work was supported by the National Science Foundation Grant No. PHY-0653623, and by the University of Arizona.

\section{Competing financial interests}
The authors declare no competing financial interests.


\end{document}